\documentclass[twocolumn,3p]{elsarticle}
\usepackage{color}

\begin{document}

\title{Optimising a balloon-borne polarimeter in the hard X-ray domain: from the PoGOLite 
Pathfinder to PoGO+}

\author[kth,okc]{M. Chauvin\corref{cor1}}
\cortext[cor1]{Corresponding author}
\ead{chauvin@kth.se}

\author[kth,okc]{M. Jackson}
\author[hun]{T. Kawano}
\author[kth,okc]{M. Kiss}
\author[kth,okc,gun]{M. Kole}
\author[kth,okc]{V. Mikhalev}
\author[kth,okc,mpi]{E. Moretti}
\author[hun]{H. Takahashi}
\author[kth,okc]{M. Pearce}

\address[kth]{KTH Royal Institute of Technology, Department of Physics, 106 91 Stockholm, Sweden}
\address[okc]{The Oskar Klein Centre for Cosmoparticle Physics, AlbaNova University Centre, 106 91 Stockholm, Sweden}
\address[hun]{Hiroshima University, Department of Physical Science, Hiroshima 739-8526, Japan}
\address[mpi]{Max Planck Institut for Physics, D-80805 Munich, Germany}
\address[gun]{University of Geneva, CH-1211 Geneva, Switzerland}

\begin{abstract}
PoGOLite is a balloon-borne hard X-ray polarimeter dedicated to the study of point sources. 
Compton scattered events are registered using an array of plastic scintillator units to determine the polarisation of 
incident X-rays in the energy range 20 - 240 keV.
In 2013, a near circumpolar balloon flight of 14 days duration was completed after launch from Esrange, Sweden, 
resulting in a measurement of the linear polarisation of the Crab emission.
Building on the experience gained from this Pathfinder flight, the polarimeter is being 
modified to improve performance for a second flight in 2016. Such optimisations, 
based on Geant4 Monte Carlo simulations, take into account the source characteristics, the 
instrument response and the background environment which is dominated by atmospheric 
neutrons.
This paper describes the optimisation 
of the polarimeter and details the associated increase in performance. The resulting design, PoGO+,
 is expected to improve the Minimum Detectable Polarisation 
(MDP) for the Crab from 19.8\% to 11.1\% for a 5 day flight. 
Assuming the same Crab polarisation fraction as measured during the 2013 flight, this improvement in MDP will allow a 5$\sigma$ constrained result. It will also allow the study of the nebula emission
only (Crab off-pulse) and Cygnus X-1 if in the hard state.
\end{abstract}


\maketitle

\section{Introduction}
Measuring the linear polarisation of X-ray emissions from astrophysical sources gives unique insight into the emission 
mechanisms at work and the geometry of the emitting region \citep{1997SSRv...82..309L,2011APh....34..550K}. 
Due to the 
difficulties in making sensitive measurements in this energy domain, only a few 
instruments have been able to successfully detect polarisation 
\citep{1976ApJ...208L.125W,2008ApJ...688L..29F,2008Sci...321.1183D,2015arXiv151102735C}. 
The delicate control of systematic effects benefits from instruments specifically designed for polarimetric measurements.

PoGOLite is a purpose-built polarimeter working in the energy range 20 - 240 keV. 
It determines the linear polarisation of hard X-ray emission from point sources by measuring the distribution of Compton 
scattering angles in a plastic scintillator detector array. 
The PoGOLite Pathfinder mission was designed to validate the instrument concept. 
Launched from the Esrange Space Centre at 08:18 UT on July 12th 2013, the payload was airborne for 14 days, making an almost 
complete circumpolar flight around the North Pole (average latitude of $68^\circ$). The prominent Crab X-ray source was 
observed for 14 hours. Data from this first flight revealed limitations of the design and a challenging background environment 
dominated by atmospheric neutrons. Based on the accumulated experience, a new design that will significantly improve the 
polarimeter performance is proposed. 

\begin{figure*}[ht!]\centering
  \includegraphics[width=\textwidth]{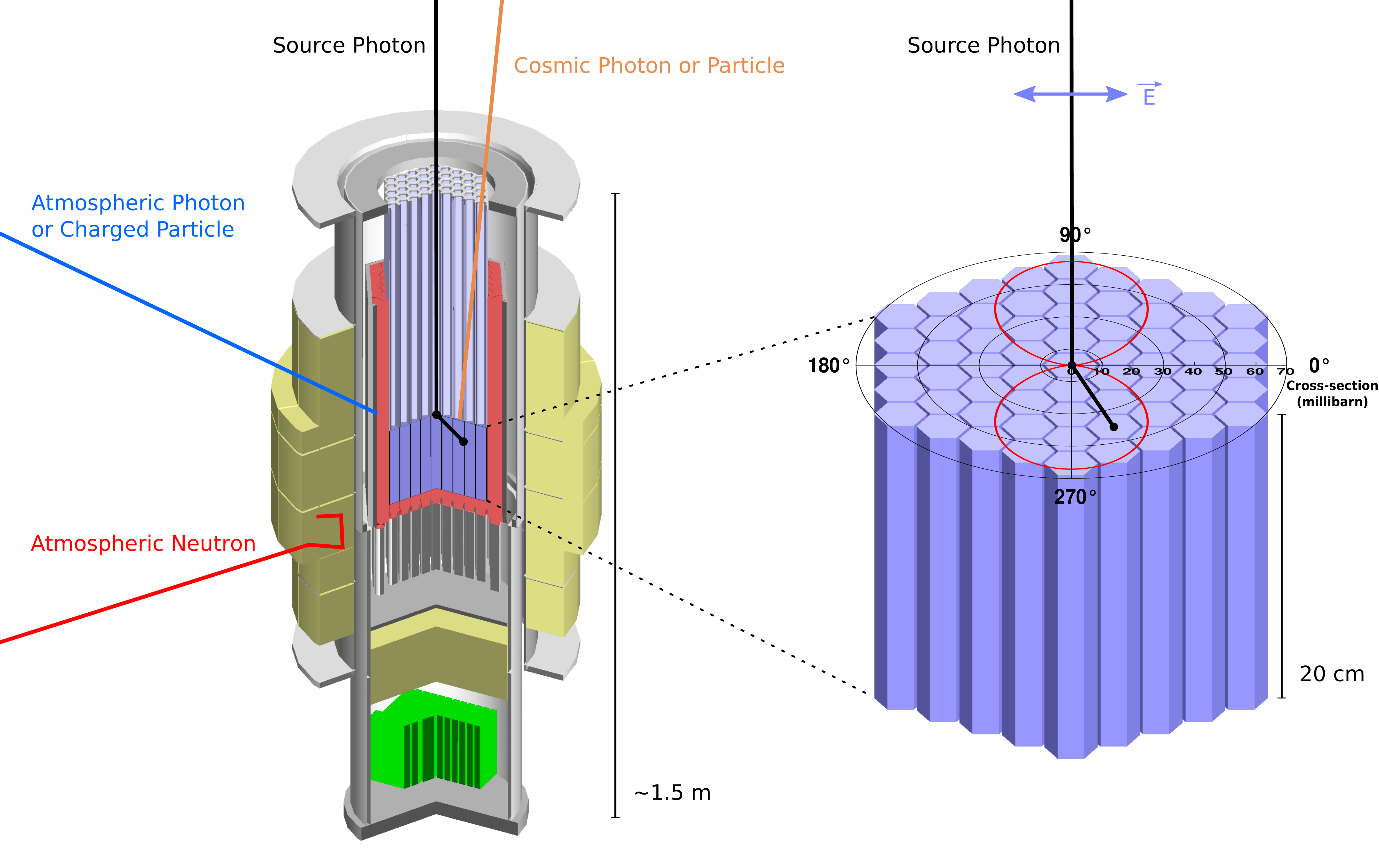}
  \caption{PoGOLite Pathfinder schematic view and detection principle. Left: the detector made of plastic scintillators (blue), 
  the BGO anti-coincidence (red), the active plastic scintillator collimators (light blue), the photomultiplier tubes (grey) 
  and the electronic components (green) are enclosed in a cylindrical housing. This inner housing is rotated around the 
  viewing axis in order to remove systematic effects in the measurements. 
The detectors are surrounded by a passive shield made of polyethylene (yellow) to reduce the
neutron background. The background components and their interactions in the instrument are 
represented by different coloured lines. Right: close-up view of the detector assembly. 
The 61 plastic scintillator rods, about 3 cm wide and 20 cm long,
are closely packed
to provide the azimuthal angle between two consecutive interactions (black dots). The Compton 
cross-section is overlayed on top (not to scale) to show the preferential scattering direction of the incoming 
polarised photons.}
  \label{Fig1}
\end{figure*}

After describing the instrument and its numerical simulation, the different modifications are presented along with
their associated increase in polarimetric performance in terms of Minimum Detectable Polarisation 
(MDP)~\citep{2010SPIE.7732E..0EW}. 

In the final section the overall 
performance of the new design, PoGO+, is discussed for the Crab and \mbox{Cygnus X-1} during the next balloon flight, planned for the summer of 2016.

\section{The PoGOLite instrument}
PoGOLite is a polarimeter making use of Compton scattering kinematics. When 
polarised photons undergo Compton scattering they have higher
probability to scatter perpendicular to their polarisation vector. This is described
by the Klein-Nishina differential cross-section:
\begin{equation}
\frac{d\sigma}{d\Omega}=\frac{1}{2}r_0^2\epsilon^2 \left[\epsilon+\epsilon^{-1}-\sin^2\theta\cos^2\phi \right]
\label{KN_equ}
\end{equation}
where $r_0$ is the classical electron radius, $\epsilon$ is the ratio between the scattered 
and incident photon energies, $\theta$ is the polar scattering angle and $\phi$ is the azimuthal scattering angle defined as 
the angle to the electric field vector. Within a polarimeter, 
this anisotropic process
causes a modulation in the detected azimuthal angles and measuring the phase and amplitude of this modulation allows the 
polarisation angle and polarisation fraction of the source flux to be determined.

\begin{figure*}[ht!]\centering
  \includegraphics[width=0.9\textwidth]{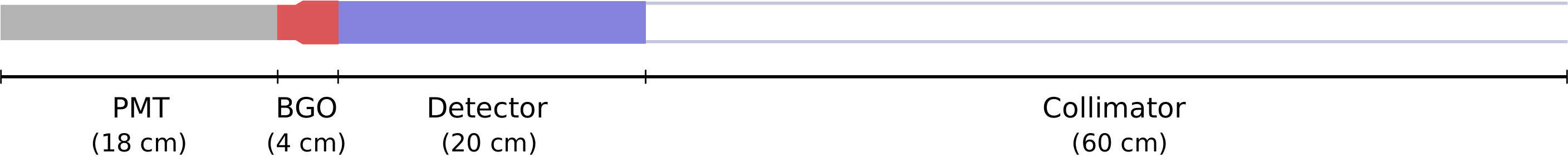}
  \caption{Sketch of a PoGOLite Pathfinder detector unit. The main detector (blue) is sandwiched between a BGO crystal (red) 
  and a hollow 2~mm thick active collimator made of plastic scintillator (light blue). The three scintillators are read-out by 
  the same PMT (grey) and their signals are identified by their different waveform shape.}
  \label{Fig2}
\end{figure*}

The PoGOLite Pathfinder uses an array of 61 plastic scintillator rods, giving an exposed detector 
area of 298 cm$^{2}$, and providing a high cross-section 
for scattering X-ray photons. The hexagonal rods are 
closely-packed to provide the azimuthal scattering angle of a photon interacting twice in 
the detector. Following the detection of an energy deposit above $\sim$20 keV in 
one detector cell (assumed to be a photoelectric absorption event), the remaining cells are 
checked for a coincident lower energy deposit above $\sim$0.5 keV (assumed to be a Compton scattering event). 
Each detector cell has a dynamic range of 20 - 120 keV. Since two-hit events are used for the 
polarisation analysis, the maximum (theoretical) energy range of the instrument becomes 20 - 240 keV.
The polarimeter and detection concept are illustrated in Figure~\ref{Fig1}.

A detector rod (20 cm long) is sandwiched between two anti-coincidence components: a BGO scintillator (4 cm long) and an 
active collimator (60 cm long) made of plastic scintillator (see Figure~\ref{Fig2}). 
The three stacked scintillators are read out by the 
same photomultiplier tube (PMT) - based on the Hamamatsu R7899 design but 
modified to reduce the dark current. The sandwiched detector has a faster rise time to 
allow event discrimination based on the pulse shape. 
The anti-coincidence is complemented by 30 rods 
of BGO scintillators (60 cm tall) placed around the main detector cells. 
PMT signals are fed to charge-sensitive 
amplifiers, the outputs of which are sampled with 12 bit precision by flash analog-to-digital 
converters (ADC) operating at 37.5 MHz. Candidate polarisation events are identified through 
recorded energy deposits. An initial rejection of background events occurs using ADC waveform 
information and signals from the BGO veto system. The remaining waveform data is saved for 
post-flight analysis. The assembly housing the detector and anti-coincidence system is rotated during observations to 
eliminate possible systematic effects, e.g. due to differences in response between detector 
cells, as well as to provide a continuous distribution of scattering angles. 
A passive neutron shield made of polyethylene additionally surrounds the 
instrument. A detailed description of the instrument and data acquisition system 
can be found in~\cite{2015ExA...tmp...55C}.

\section{The PoGOLite simulation}\label{Sim}
\begin{figure*}[ht!]\centering
  \includegraphics[width=\textwidth]{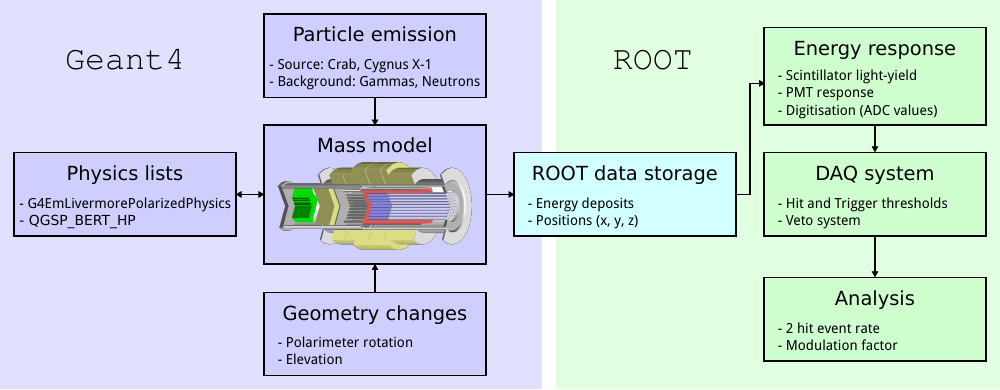}
  \caption{PoGOLite simulation flow-chart. The simulation is divided in two main parts, one simulating the particle interactions (using the Geant4 toolkit, blue) and one reproducing the instrument response (using the ROOT environment, green).}
  \label{Fig3}
\end{figure*}

The PoGOLite simulation allows the performance of the polarimeter to be studied, depending on the observation environment and detector configuration. 
The simulation is divided in two parts (see Figure \ref{Fig3}). 
The first part simulates particle interactions in the instrument and includes source emission (e.g. Crab) and background processes (neutrons and photons).
The second part reproduces the instrument response (scintillator light-yield, PMT behaviour and data acquisition system).

The first part of the simulation has been developed using Geant4~\citep{2003NIMPA.506..250A} 10.0 patch-02 and includes the 
instrument mass model (Figure~\ref{Fig1}). As Geant4 is designed to simulate static geometries, a custom function has been 
developed to allow the rotation of the polarimeter. Particles are generated according to source and background emission 
models including air 
density depending on the instrument altitude and source elevation. The particle 
interactions are simulated using appropriate standard physics lists, G4EmLivermorePolarizedPhysics for photons 
and QGSP\_BERT\_HP for neutrons.
The energy and position of each interaction is stored in a ROOT~\citep{1997NIMPA.389...81B} tree format.

The second part of the simulation 
reproduces the conversion of energy deposits into ADC values and emulates the data acquisition system of PoGOLite. Every 
interaction is converted into scintillation light, then into a PMT signal~\cite{2009NIMPA.600..609M}. The different scintillator material 
properties, their light-yield non-linearity (quenching), the scintillation light attenuation, the production of 
photo-electrons, the multiplication of charges and the conversion into ADC values (digitisation) are all accounted for. Every detector 
cell is independently treated using measured calibration parameters such as the conversion from energy to number of photo-electrons 
relation (scintillator-dependent) and the single photo-electron peak position in ADC values (PMT-dependent). Unintentional optical 
cross-talk between detector cells (discovered after the 2013 flight) is also taken into account based on laboratory 
measurements performed with a pulsed blue LED. 
Once deposited energies from Geant4 are converted into ADC values, event selection criteria imposed by the data acquisition 
system of PoGOLite are applied. First, every detector cell yielding a PMT signal amplitude less than 10 ADC values is ignored 
(zero-suppression hit threshold). 
Then a trigger is issued if one of the signal amplitude exceeds 300 ADC values.
After applying these thresholds, the veto system rejects the event if there is any signal in the anti-coincidence 
system or if any energy deposit exceeds an upper discriminator threshold (i.e. likely corresponding to a cosmic ray 
interaction).
For interactions in any one of the 30 BGO units that surround the main detector, the event is discarded if above the hit 
threshold. For the active collimators and the BGO scintillators placed above and below the detector respectively, the 
discrimination is based on the waveform shape. 
Using laboratory measurements made on single scintillator elements 
separately coupled to a PMT, analytical functions have been established linking the energy deposit to 
the waveform amplitude after two typical rise times called ``fast output'' ($\sim$107 ns) and ``slow output'' ($\sim$400 ns). 
These functions allow the veto system based on waveform discrimination to be implemented in the 
simulation using the same algorithm 
as on-board PoGOLite. Only events that pass the veto with two interactions in different detector 
cells are considered as polarisation events. The azimuthal angle is determined and filled in a histogram resulting in 
a sinusoidal modulation curve from which the polarisation angle (phase) and polarisation fraction (amplitude) are extracted.

The simulation has been developed and validated using data taken during the preflight calibration of the 
polarimeter~\cite{2016APh....72....1C}. The modulation factor for a 100\% polarised source ($M_{100}$) of 53.3 keV 
photons has been found to be $(21.3 \pm 0.9)\%$ from measurement and $(23.7 \pm 0.14)\%$ from simulation. These 
simulation results were based on a simplified implementation of the veto system, which assumed that discrimination between a 
waveform from the detector and one from the anti-coincidence was not possible below 100 ADC values (∼5 keV). After updating the 
simulation with a more realistic implementation of the veto system as described above, $M_{100}$ is $(22.97 \pm 0.22)\%$, 
which is compatible with measurements within 1.5 standard deviations. 

\begin{figure*}[ht!]\centering
  \includegraphics[width=\textwidth]{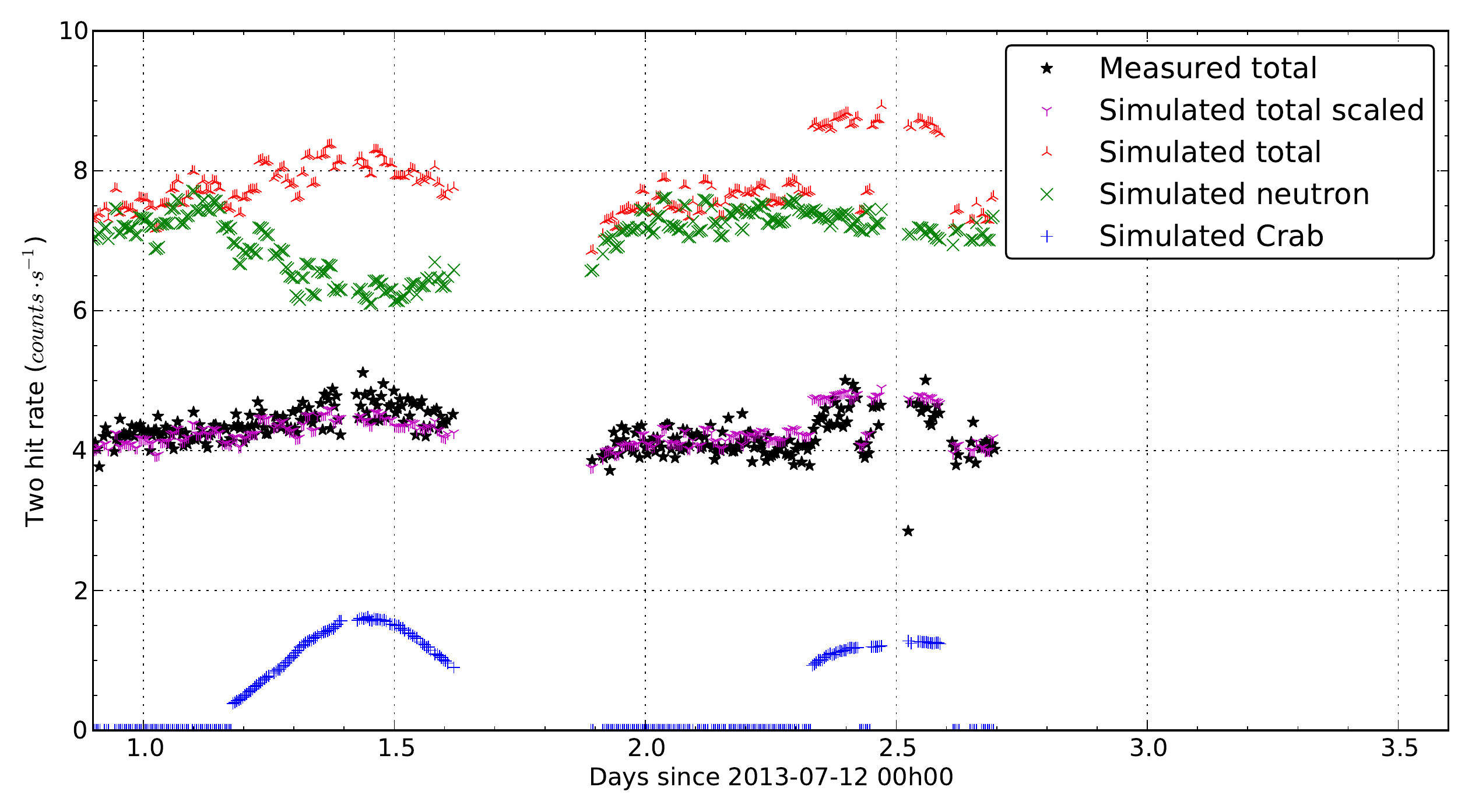}
  \caption{PoGOLite Pathfinder two hit (polarisation event) rate during the 2013 flight. Measured rate is shown (black stars)
with the simulated rates; Crab (blue pluses), neutron (green crosses), total (red tripods up)
and scaled total (purple tripods down). The total includes the Crab, the neutron and the 
photon background rates. The latter is not shown and contributes as a constant 
$0.28\ \mathrm{counts}{\cdot}\mathrm{s}^{-1}$.
The simulation is scaled down to account for the waveform selection efficiency of the instrument data. 
The gaps in the data correspond to intervals where the data acquisition was not running.
}
  \label{Fig4}
\end{figure*}

Waveform discrimination for the 2013 flight data was found to be more challenging than for the preflight calibrations. There 
are two reasons for this. Firstly, the variable temperature environment during the flight affect the waveform pedestal level so a 
temperature correction needed to be applied.
Secondly, many waveforms with a shape corresponding to a mix of ``slow'' rise-time from the collimator and ``fast'' rise-time 
from the detector were found in the data. These super-imposed waveforms, coming from simultaneous energy deposits in a 
collimator and a detector of the same unit (e.g. via Compton scattering), are very similar to waveforms from the detector only, 
making them difficult to reject.
As a consequence, the post-flight event identification using only two variables for waveform discrimination (the ``fast output'' 
and the ``slow output'') showed poor efficiency, resulting in a low signal-to-background ratio.

In order to improve the veto efficiency, a new discrimination technique using a principal component analysis (PCA) was 
developed and used on the flight data~\cite{2015ExA...tmp...55C}. This method, making use of more information from the 
waveform shape (ten variables instead of two), reduces the total number of accepted two-hit events 
(from 8.4 to 4.6 Hz) but significantly increases the signal-to-background ratio 
(from 1:7 to 1:4).
Figure~\ref{Fig4} shows the two-hit polarisation event count rate observed during the 2013 flight, showing a clear increase 
during the two Crab observations. The simulated Crab rate, simulated 
neutron rate and simulated total rate are also shown. The photon background, composed of the cosmic X-ray background, 
secondary atmospheric photons and the 511 keV line, is included in the total but is not shown as it contributes to only 
$0.28\ \mathrm{counts}{\cdot}\mathrm{s}^{-1}$ and is assumed to be constant. 
The simulated Crab rate takes into account the altitude and pointing elevation of the polarimeter. 
The neutron rate is simulated using the model 
presented in~\citep{2015APh....62..230K} and takes into account the 
directionality of the incident neutron flux, variations in altitude and 
magnetic latitude of the instrument as well as the solar activity at the time 
of the measurement. The reduction in neutron rate around 1.15 days, as shown 
in Figure 4, is confirmed by a dedicated on-board neutron detector and coincides 
precisely with a decrease in the neutron flux measured by the Oulu Neutron Monitor\footnote{On-ground 
cosmic ray station, http://cosmicrays.oulu.fi/.}. Additional details regarding the 
flight are presented in~\cite{2015ExA...tmp...55C}. 
The simulated total rate is then scaled down to account for the PCA waveform selection efficiency. In the next sections, this 
scaling is applied to the simulations to be representative of the actual data analysis.

\section{New design: PoGO+}
A wealth of improvements to the PoGOLite Pathfinder polarimeter design have been identified based on the analysis of the 2013 
flight data. This section reports on the performance study of different configurations to optimise the polarimeter for the 
next flight. These studies, made using the PoGOLite simulation, are presented in an incremental (cumulative) way to show the 
overall increase in performance. The incremental design changes are likely not independent and therefore have no realistic 
interpretation when treated separately. To be representative of a real flight, the Crab simulations assume the average 
column density as experienced during the 2013 flight, and the neutron background simulations assume the average neutron activity 
that occurred during the flight (based on the model from~\citep{2015APh....62..230K}). 
The Minimum Detectable Polarisation at 99\% confidence level (MDP) is used as a figure of merit to compare 
the different configurations. It is defined as:
\begin{equation}
  \mathrm{MDP}=\frac{4.29}{M_{100} R_\mathrm{S}}\sqrt{\frac{R_\mathrm{S}+R_\mathrm{B}}{T}}
\end{equation}
where $R_\mathrm{S}$ is the source rate, $R_\mathrm{B}$ the background rate and $T$ the observation time. For all following 
studies, the observation time is set to 6 hours (daily Crab observation window).

\subsection{Scintillator coating}
\begin{figure}[ht!]\centering
  \includegraphics[width=0.48\textwidth]{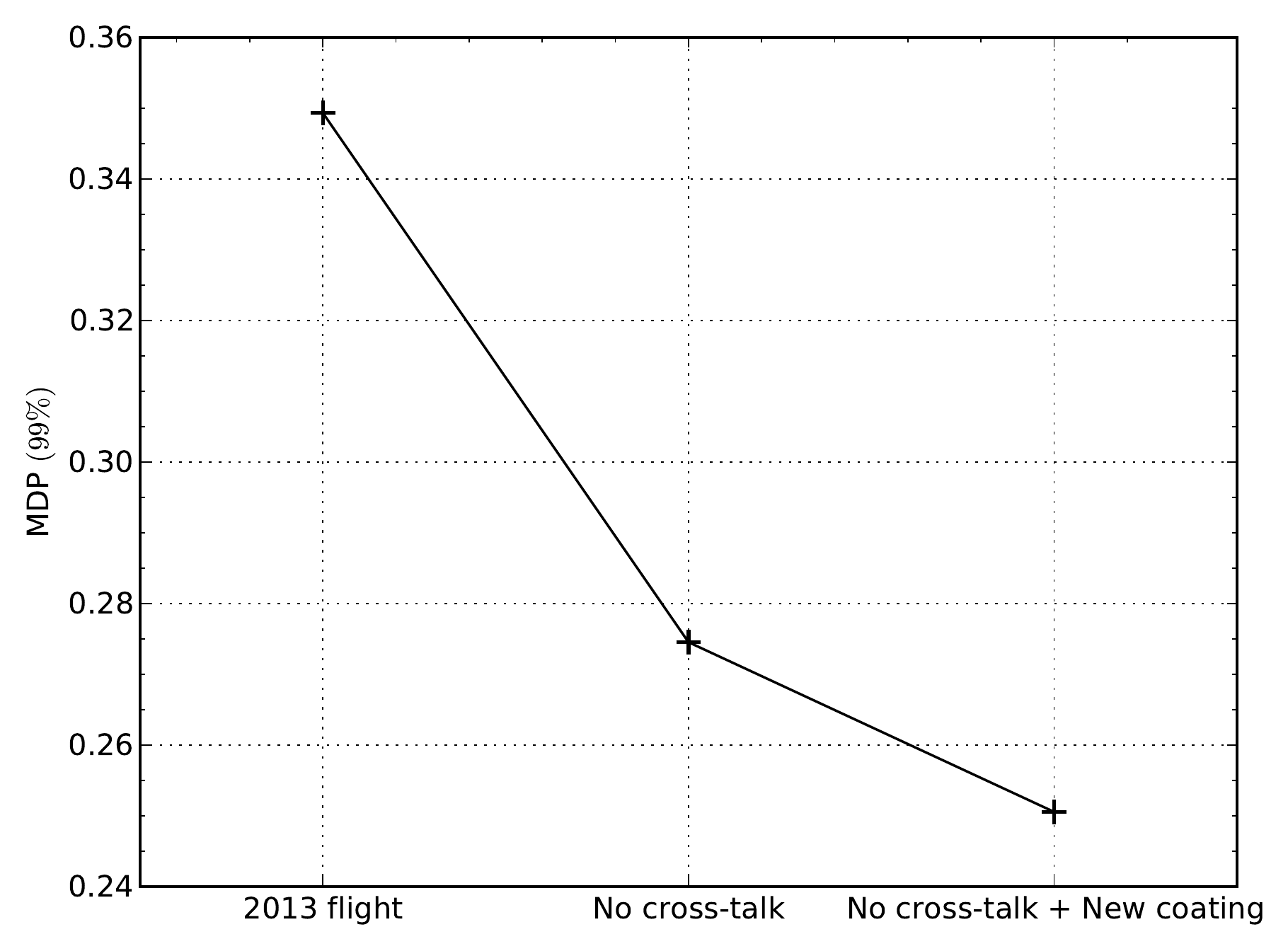}
  \caption{Polarimeter performance in terms of MDP for three configurations. The left-most point corresponds to the original 
  design that was launched in 2013, the centre point to the suppression of optical cross-talk and the right-most point to 
  additionally changing the reflective coating of the detector cells.}
  \label{Fig5}
\end{figure}

The relatively low $M_{100}$ of the PoGOLite Pathfinder has been found to be due to unintentional leakage of scintillation 
light between detector cells. 
A one-hit event causing a leakage can incorrectly be tagged as a two-hit polarisation event 
("false positive") and a two-hit polarisation event with leakage can erroneously be rejected 
as having higher multiplicity ("false negative"). Although the fraction of light leaking into other detector units 
is small ($\sim$1\%), 
it significantly degrades the polarimeter performance reducing the $M_{100}$ by 
$\sim$27\% (relative).
Figure~\ref{Fig5} shows the improvement in 
terms of MDP when the optical cross-talk is suppressed. 
This is achieved by changing the 
wrapping of the detector cells, using opaque DuPont Tedlar sheets and black heat-shrink material.
The light collection can also be improved which directly impacts the polarimeter 
performance by decreasing the lower limit of the energy range and increasing the $M_{100}$. 
This is of crucial importance for Compton polarimeters such as PoGOLite, where low 
energy deposits must be clearly distinguishable\footnote{\mbox{25 keV} photon deposits \mbox{$\sim$1.2 keV} when Compton 
scattering through \mbox{90$^{\circ}$}.}, since the Crab has an emission spectrum 
${\sim}9.7(E/\mathrm{keV)^{-2.1}photons{\cdot}cm^{-2}{\cdot}s^{-1}{\cdot}keV^{-1}}$. By changing the reflective coating using 
white PTFE tape instead of $\mathrm{BaSO}_4$-loaded epoxy and 3M ESR Vikuiti instead of 3M VM2000 specular reflector films, 
the light collection increases by 46\%. This measured new light yield is used as an input to the simulation and the MDP 
decreases from $27.5\%$ to $25\%$ (see Figure~\ref{Fig5}). 

\subsection{Collimators}
\begin{figure}[ht!]\centering
  \includegraphics[width=0.48\textwidth]{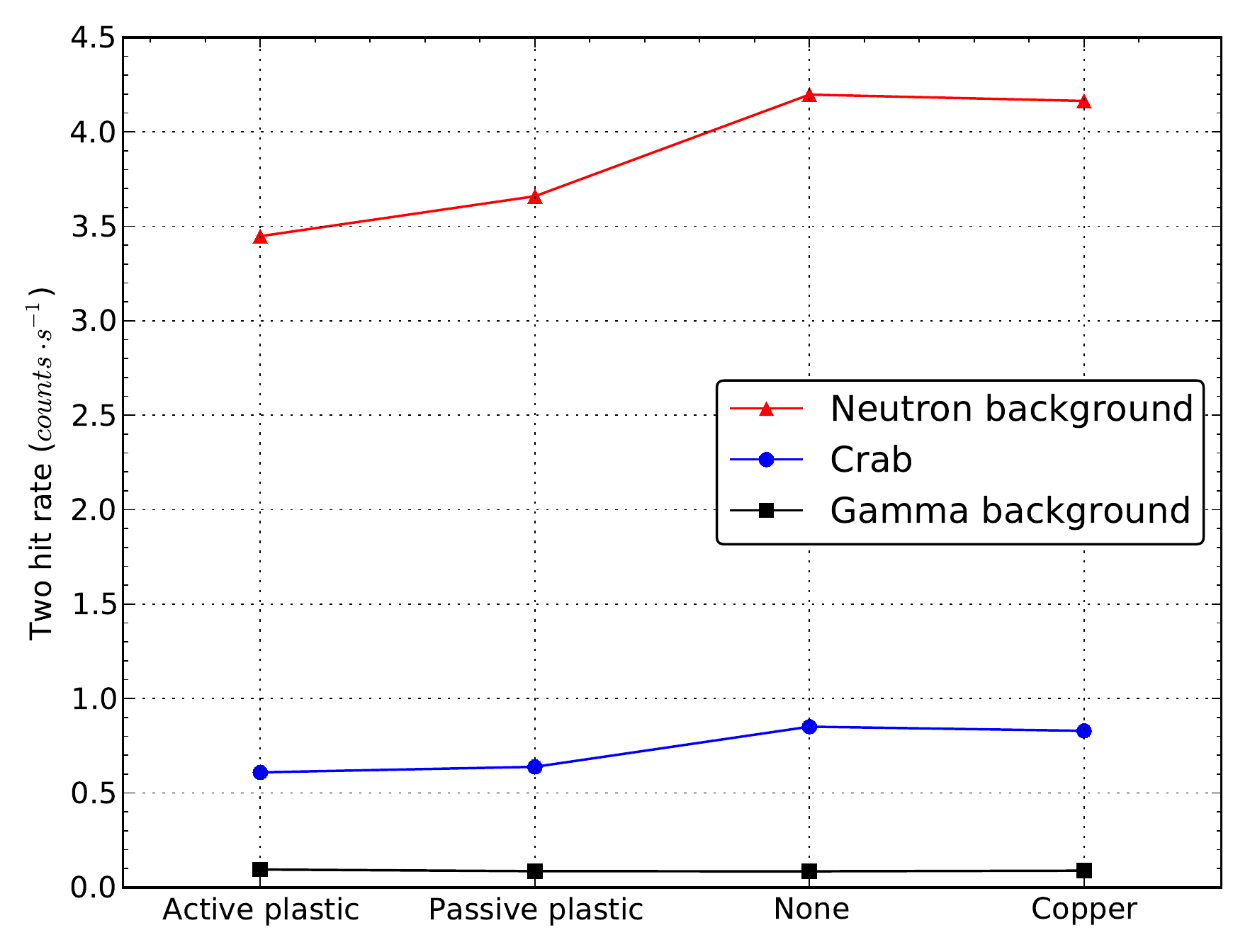}
  \includegraphics[width=0.48\textwidth]{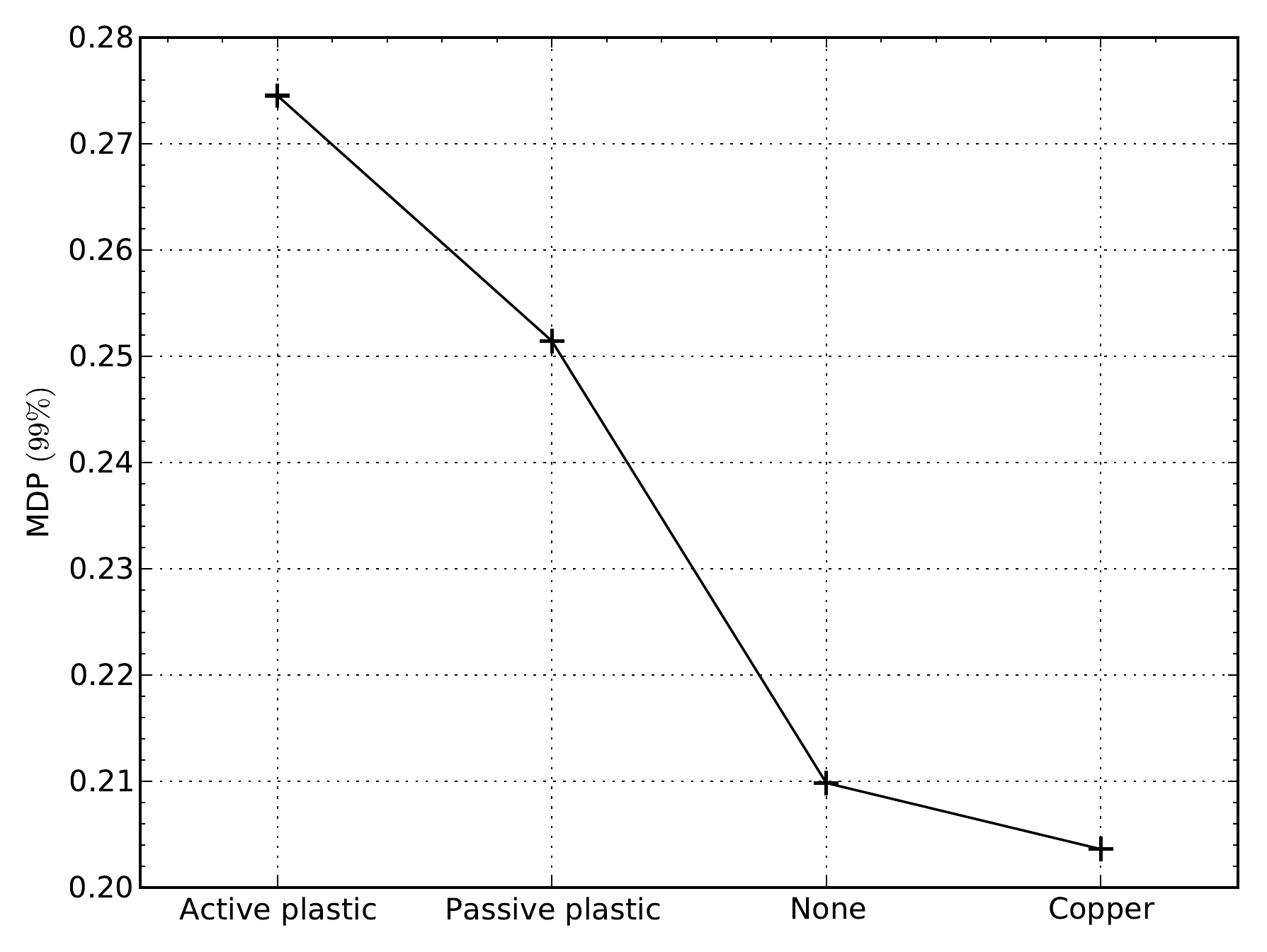}
  \caption{Polarimeter performance for four collimator types: 
``Active plastic'' is the original active plastic collimators wrapped in \mbox{50 $\mu$m} thick lead and tin foils, ``Passive plastic'' is the same as the original but with passive plastic (not scintillating), ``None'' is with only the \mbox{50 $\mu$m} thick lead and tin foils and ``Copper'' is one with 0.5 mm thick copper wrapped in \mbox{100 $\mu$m} thick lead and tin foils.
The top figure displays the two-hit rate of the Crab (blue dots), 
the photon background (black squares) and the neutron background (red triangles) for each configuration. 
The bottom figure shows the corresponding performance in terms of MDP.}
  \label{Fig6}
\end{figure}
The original design of the PoGOLite Pathfinder includes active collimators made of plastic scintillators read out by a PMT
after the scintillation light traverses the main plastic detector and a BGO scintillator (Figure \ref{Fig2}). 
Since the plastic collimators are long and have a 2~mm wall, the loss of 
scintillation light for interactions happening far from the PMT results in a decrease in the detected signal amplitude.
In general, for high amplitude signals, it is easy to distinguish in which component the interaction has taken place. This, however, 
becomes increasingly difficult for low energies. Tests 
performed on ground with a radioactive source illuminating only the collimator showed good performance above the trigger threshold
\mbox{($\sim$20 keV)} with $93\%$ rejection efficiency but rather poor performance at lower energies 
(between the hit threshold $\sim$0.5 keV and the trigger threshold) with $20\%$ 
rejection efficiency \citep{2016APh....72....1C}. 
In addition, as mentioned in Section \ref{Sim}, results from the 2013 flight showed waveform discrimination limitations when 
simultaneous interactions occur in the collimator and the detector of the same unit. 
In light of this, several passive collimators were studied to replace the original active collimators of the PoGOLite 
Pathfinder. 
Figure~\ref{Fig6} shows the two-hit rate of the Crab, the neutron background and the photon background for four configurations, 
as well as the corresponding performance in terms of MDP. 
The second configuration (``Passive plastic'') leads to a better MDP than the original active collimator (``Active plastic'') 
thanks to the improved reflective coating, which improves both the Crab two-hit rate 
(+4.8\% relative)
and the $M_{100}$ 
(+7.0\% relative).
The first configuration does not benefit from the improved coating because active collimators need optical coupling to the 
detectors to be read-out by a single PMT (scintillation photons produced in the detector can escape to the collimator). 
The third configuration is a purely hypothetical scenario, where the collimator material itself has been removed but the 
protective lead and tin foils are still in place. 
The absence of the collimator material 
eliminates the shadowing of the underlying detectors, resulting in an increase in the exposed 
detector area (+36.6\% relative) thus increasing the source rate. The 
$M_{100}$ is also increased (+4.1\% relative) for the same reasons as the second 
configuration. Although providing less shielding, the increase in source rate and $M_{100}$ 
leads to a better MDP. 
However, this 
scenario is difficult to realise, as some supporting structure is needed for holding the lead and tin foils in place. 
In the fourth configuration, the copper 
collimators also increase the exposed detector area (+26.8\% relative) and the $M_{100}$ (+9.6\% relative) 
compared to configurations 1 and 2, and provide better shielding than configuration 3 leading to a better MDP.
The final configuration, with the active plastic collimator replaced by thin-walled copper collimators wrapped in lead and 
tin foils, will therefore be adopted for PoGO+, decreasing the MDP to 20.4\%.

\subsection{Detector length}
\begin{figure}[ht!]\centering
  \includegraphics[width=0.48\textwidth]{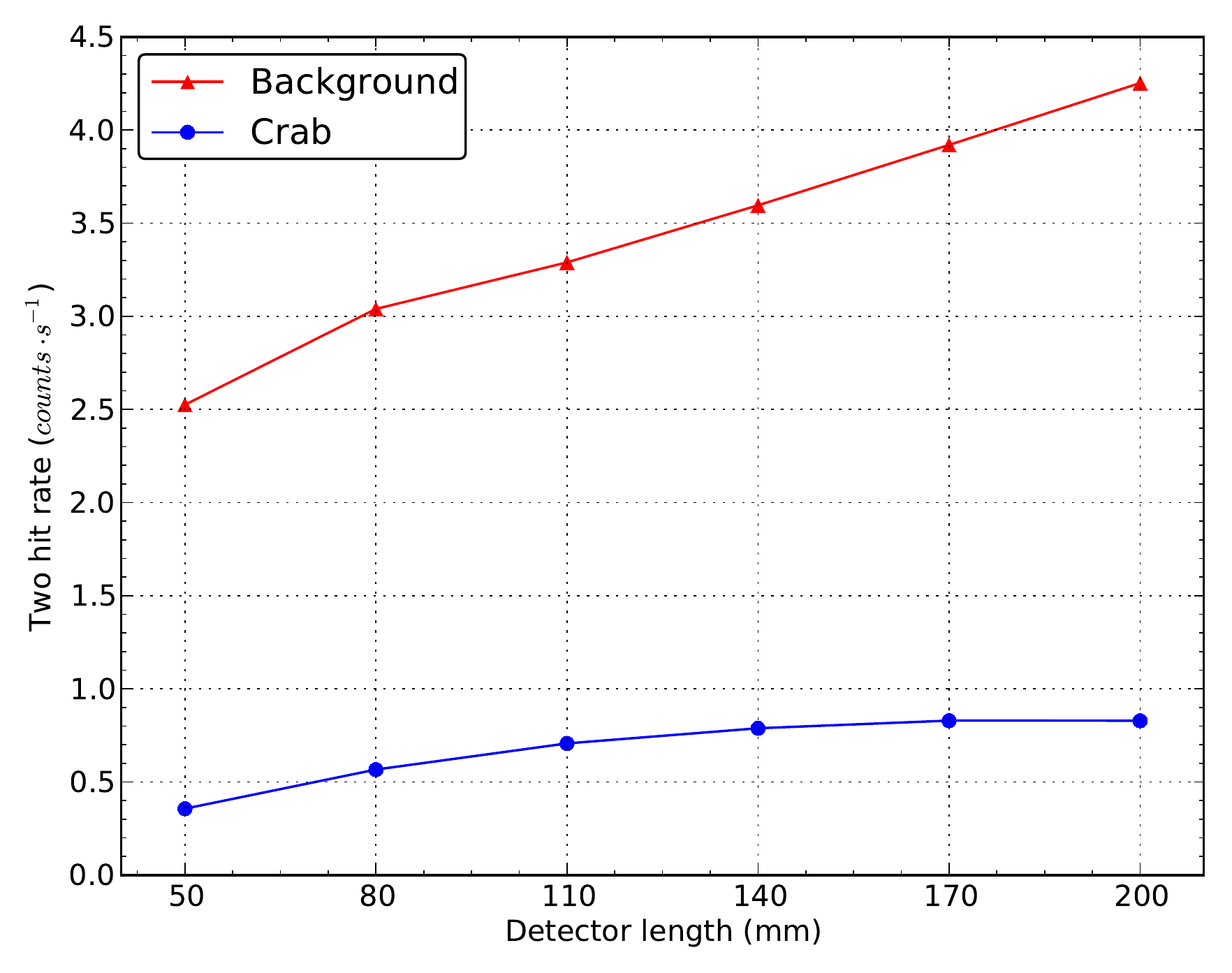}
  \includegraphics[width=0.48\textwidth]{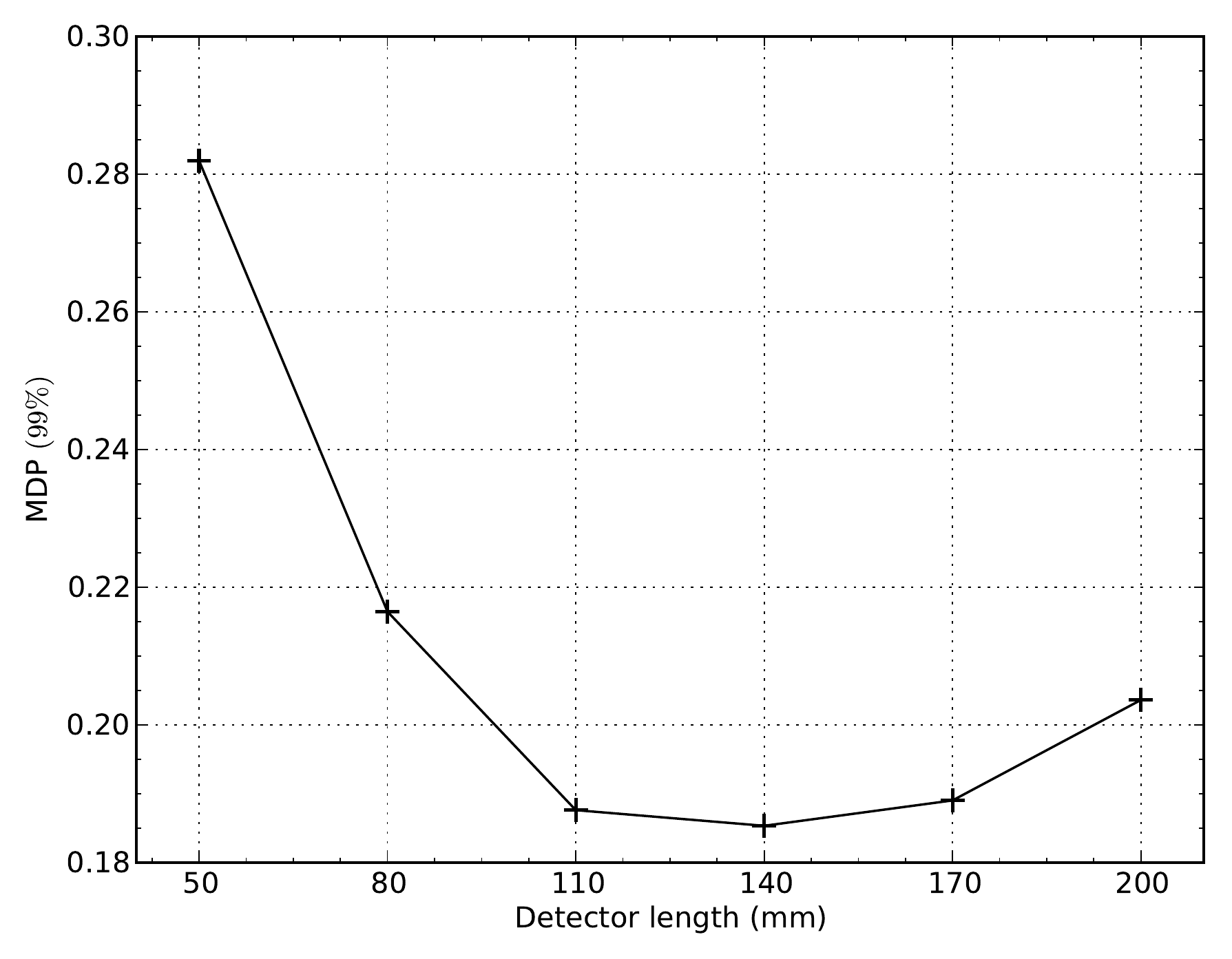}
  \caption{Polarimeter performance for varying detector lengths. The top plot shows the two-hit rate of the Crab (blue dots) and the background rate (red triangles) for each configuration, while the bottom figure shows the corresponding performance in terms of MDP.}
  \label{Fig7}
\end{figure}
Another way to improve the signal-to-background ratio is to optimize the size of the detectors. As the background is
coming from every direction (proportional to the detector volume) whereas the source flux only comes from within the 
aperture (proportional to the detector area), having long detector rods enhances sensitivity to background. To study this 
effect, the length of the main plastic scintillator detectors has been decreased iteratively in Geant4, for the range 200~mm 
down to 50~mm. For each length, both Crab and background simulations are conducted. Figure~\ref{Fig7} displays the results as 
a function of the detector length, both in terms of the resulting event rate and in terms of MDP. These results include the 
change in light collection interpolated from measurements, with a shorter detector resulting in higher light-collection than a 
longer one\footnote{Comparative tests have been conducted with \mbox{10 cm} and \mbox{20 cm} long scintillator pieces, 
allowing a linear interpolation to be used.}. The background rate increases almost linearly with the detector length while the 
increase in the Crab rate saturates above \mbox{110 mm}. 
A lower 
total rate will further decrease the MDP by increasing the live-time of the 
instrument\footnote{The live-time is not known at the moment since PoGO+ 
will use new electronics.}, therefore privileging shorter detectors. As the MDP 
is found to be stable between 120 mm and 140 mm, it has been decided that 120 mm long 
detectors will be adopted for PoGO+, providing better live-time as well as better light 
collection.
The increase in performance resulting from this length 
optimisation is from 20.4\% MDP to 18.6\% MDP.

\subsection{Neutron shield}
\begin{figure*}[ht!]\centering
  [a]{\includegraphics[width=0.4\textwidth]{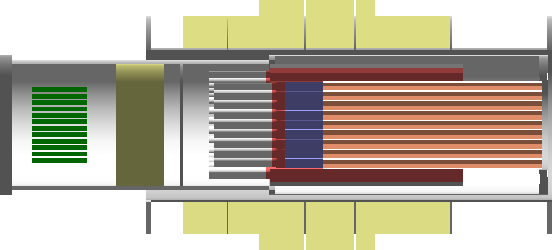}}\hspace{1cm}
  [b]{\includegraphics[width=0.4\textwidth]{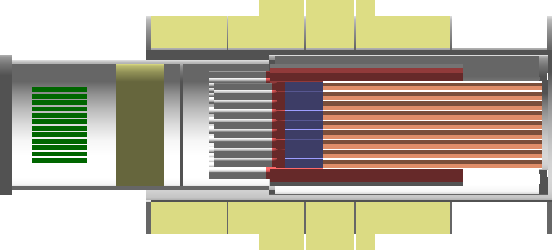}}\vspace{0.3cm}
  [c]{\includegraphics[width=0.4\textwidth]{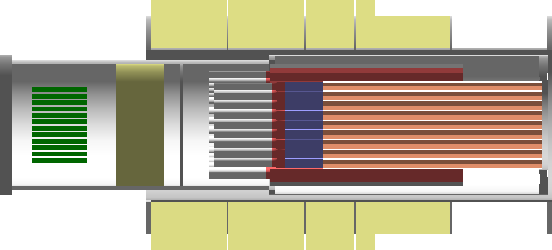}}\hspace{1cm}
  [d]{\includegraphics[width=0.4\textwidth]{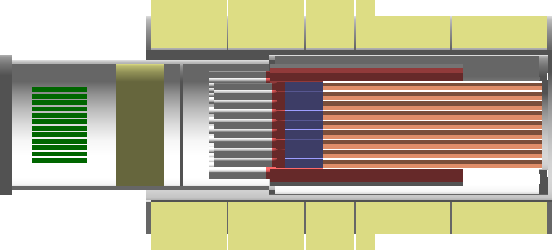}}\vspace{0.1cm}
  \caption{The different simulated shield configurations used for studying the neutron background. The PoGOLite Pathfinder 
original shield (a), one with additional 10 cm thick polyethylene at the bottom (b), one with all the bottom
part thickness increased to 15 cm (c) and one with both 15 cm thick at the bottom and 10 cm thick at the top 
(d).}
  \label{Fig8}
\end{figure*}
\begin{figure}[ht!]\centering
  \includegraphics[width=0.48\textwidth]{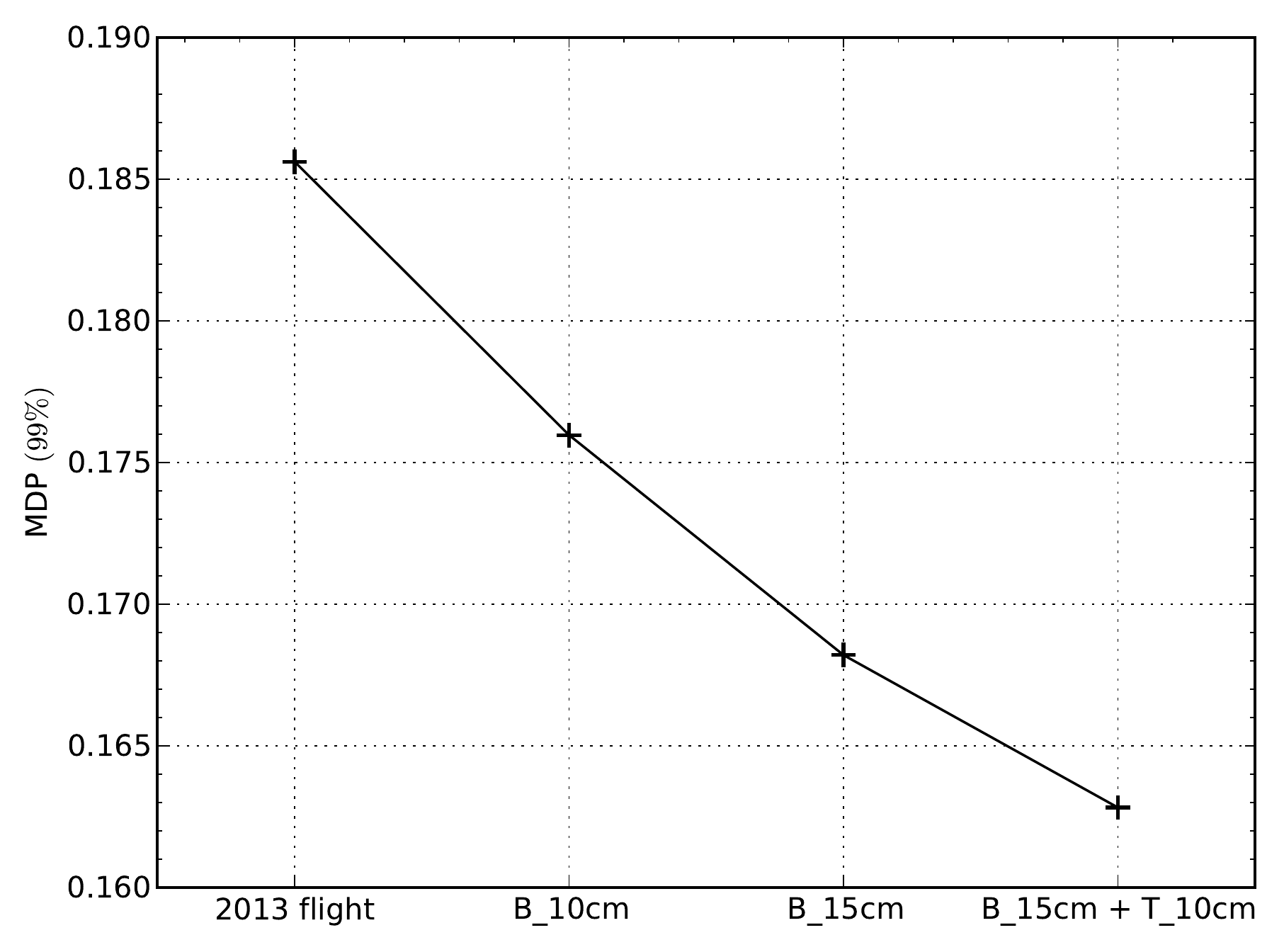}
  \caption{Polarimeter performance for different neutron shield configurations. Labels "2013 flight", "\mbox{B\_10 cm}", 
  "\mbox{B\_15 cm}", and "\mbox{B\_15 cm + T\_10 cm}" correspond to Figure \ref{Fig8} a, b, c and d shield configurations, 
  respectively.
  }
  \label{Fig9}
\end{figure}

The main source of background for PoGOLite is from neutrons produced in the atmosphere, as measured by the 
PoGOLino experiment \citep{2015NIMPA.770...68K} and by the neutron detector on board the PoGOLite Pathfinder~\cite{2015ExA...tmp...55C}. 
This background is mitigated using a passive shield made of a hydrogen-rich material, i.e. polyethylene. The shield 
configuration used during the 2013 flight is shown in Figure~\ref{Fig8}-a. With a better understanding of the in-flight 
neutron environment, as well as a more complete simulation, it has been possible to re-optimise the shield geometry to 
further reduce the neutron background induced in the polarimeter. 
In particular, the region between 
the side and bottom neutron shield can be improved. 
Thus, three new designs have been studied in simulations. The shield geometries that 
have been considered are shown in Figure~\ref{Fig8}, while the corresponding MDP values are presented in Figure~\ref{Fig9}.
Increased shield thickness improves the MDP, but the mass increases as well. Configurations (b), (c) and (d) add 
\mbox{16.8 kg}, \mbox{51.5 kg} and \mbox{100.7 kg}, respectively, as compared to the 2013 flight geometry. The payload mass 
($\sim2$ Tonnes) dictates the achievable flight altitude and thus the observation environment (source and background fluxes). 
As configuration (b) provides a good improvement with respect to the increase in mass\footnote{It is assumed 
that the overall mass of the payload does not change.}, this configuration has been chosen for the shield of PoGO+, resulting in an MDP of \mbox{17.6\%}.

\section{PoGO+ expected performance}
After the optimisation studies presented in the previous section, PoGO+ is currently under construction at KTH Royal Institute 
of Technology. The reflective coating and optical isolation of the detector cells have been reviewed, significantly increasing 
the light yield and $M_{100}$. The collimators have been changed from active plastic scintillators to thin-walled passive 
copper, reducing the shadowing of the detector area underneath the collimators and reducing both the induced background and 
complexity of the waveform discrimination veto. The length of the main plastic scintillator detectors has been shortened from 
\mbox{20 cm} to \mbox{12 cm}, reducing the volume for background interactions while not notably affecting efficiency 
for source detection. This shortening of the detectors also improves the light-yield and the observation live-time of the 
instrument by reducing the total event rate. The neutron shield will be upgraded by adding 10~cm thick polyethylene at the 
bottom of the instrument reducing the neutron rate by 12\%. With this new design, PoGO+ is expected to reduce the MDP 
from $34.5\%$ to $17.6\%$ for a 6 hour Crab observation. 

\begin{figure}[ht!]\centering
  \includegraphics[width=0.48\textwidth]{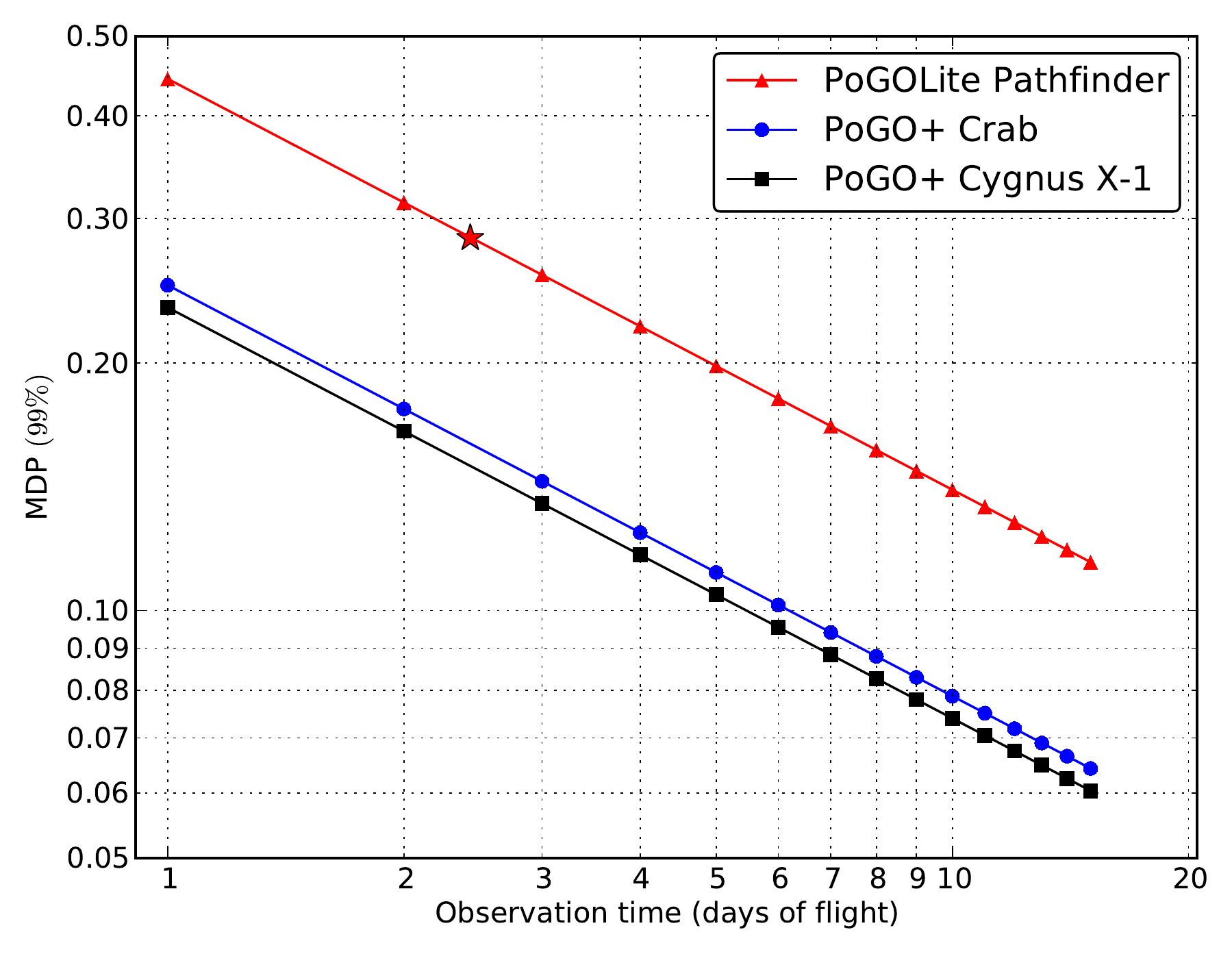}
  \caption{PoGO+ expected performance for the next flight assuming 50\% time on-source (Crab or \mbox{Cygnus X-1}) and 50\% 
  time off-source (background fields). The MDP for the Crab (blue dots) and \mbox{Cygnus X-1} (black squares) is shown as a 
  function of the number of days of flight. For the Crab an observation window of 6 hours is assumed, for \mbox{Cygnus X-1}, 
  10 hours. The PoGOLite Pathfinder performance for the Crab (red triangles) is shown as a comparison and the red star shows 
  the MDP for the Crab as observed during the 2013 flight \citep{2015arXiv151102735C}.}
  \label{Fig10}
\end{figure}

Calibration studies of the PoGOLite Pathfinder have shown that a strongly anisotropic flux coming from the side of the 
instrument can induce a fake polarisation signature~\cite{2016APh....72....1C}. It is therefore important to measure the 
background (off-source) to either show that no significant fake modulation is induced or for subtracting its modulation from 
the on-source measurement. When planning an in-flight observation, a suitable method for determining the fraction of time 
spent on-source and off-source is given by~\cite{2015APh....68...45K}:
\begin{equation}
  \frac{T_\mathrm{off}}{T_\mathrm{on}}=\frac{1}{\sqrt{1+R_\mathrm{S}/R_\mathrm{B}}}.
\end{equation}
For instruments with modest signal-to-background ratio like PoGO+ ($R_\mathrm{S}/R_\mathrm{B} \sim1/4$),
$T_\mathrm{off}/T_\mathrm{on}$ approaches unity so 
PoGO+ will observe on-source and off-source for equal duration. Simulations with off-axis 
pointing scenarios show that the field of view is $\sim$2 degrees, defined as twice the angle 
outside which the effective area drops below 50\%. For an off-axis pointing of 5 degrees, the 
effective area is 0.07\%, making this suitable for background measurements. 
Figure~\ref{Fig10} shows the expected performance of PoGO+ assuming this 50\% on-source observation strategy. 
The MDP for the Crab is calculated assuming an observation window of 6 hours per day, meaning 3 
hours on-source and 3 hours off-source. The aforementioned MDP of $17.6\%$ for 6 hours of Crab observation 
is therefore achieved after a two-day flight. The MDP for \mbox{Cygnus X-1} (black squares) is calculated assuming 
that the source is in the high state\footnote{A simple power law 
$2(E/\mathrm{keV)^{-1.7}photons{\cdot}cm^{-2}{\cdot}s^{-1}{\cdot}keV^{-1}}$ is used to represent the hard state emission of Cygnus X-1 
from 10 keV to 150 keV.} and an observation window of 10 hours per day (\mbox{Cygnus X-1} is higher in the sky than the Crab 
as seen from Esrange during the Summer). Both the Crab and \mbox{Cygnus X-1} are simulated using their respective average column density from the 2013 flight elevation and altitude profiles.
Although the Crab emission rate is higher in the PoGO+ energy range, the difference in observation conditions leads to a 
better MDP for \mbox{Cygnus X-1}. The MDP of the PoGOLite Pathfinder design and 2013 flight for the Crab are also shown for 
comparison. The expected performance of PoGO+ for a 5 day flight reaches $11.1\%$ MDP, while the corresponding number from the 
PoGOLite Pathfinder, extrapolated from the observed in-flight performance, would have been $19.8\%$ MDP. 

\begin{figure}[ht!]\centering
  [a]{\includegraphics[width=0.47\textwidth]{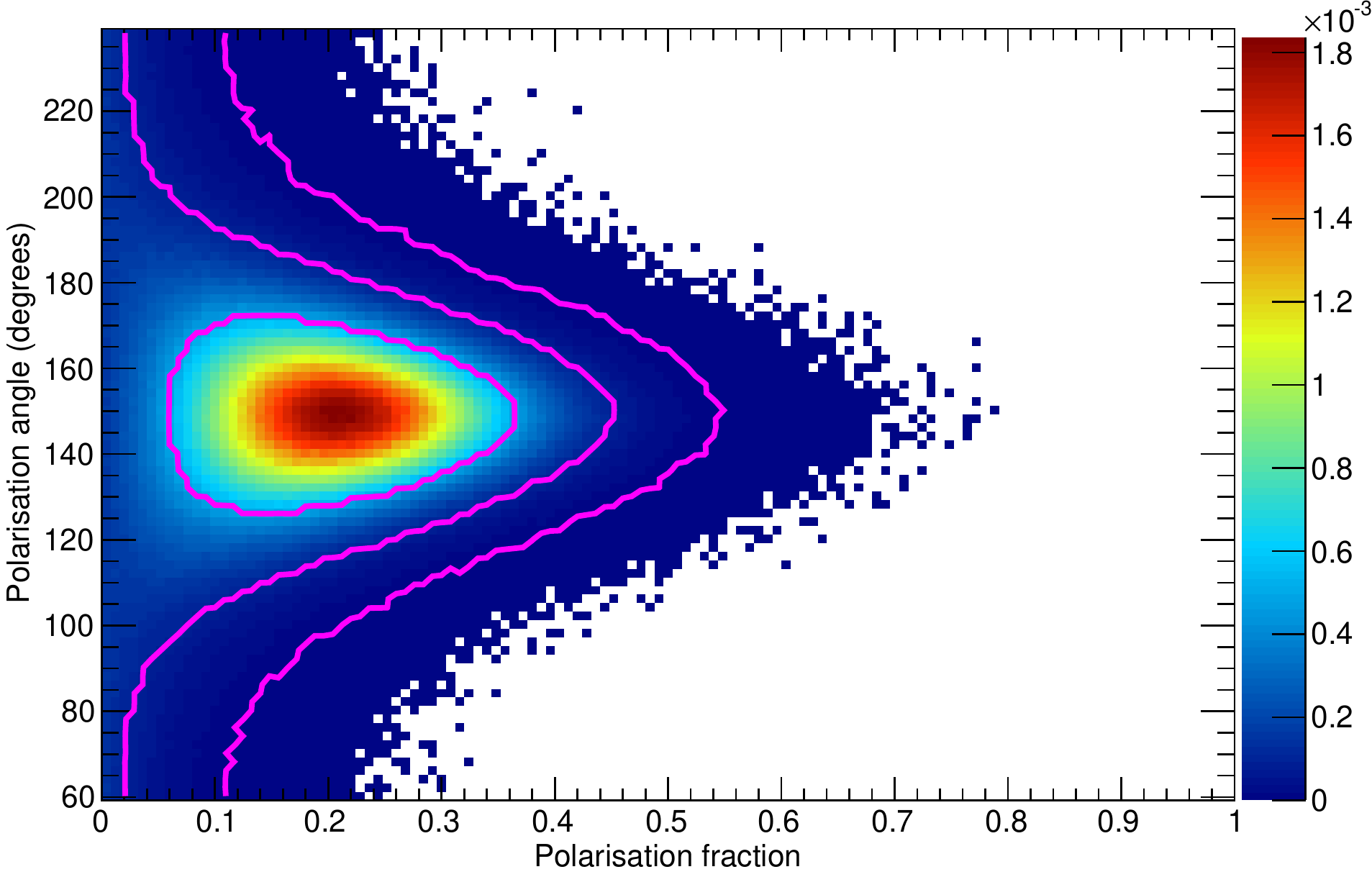}}
  [b]{\includegraphics[width=0.47\textwidth]{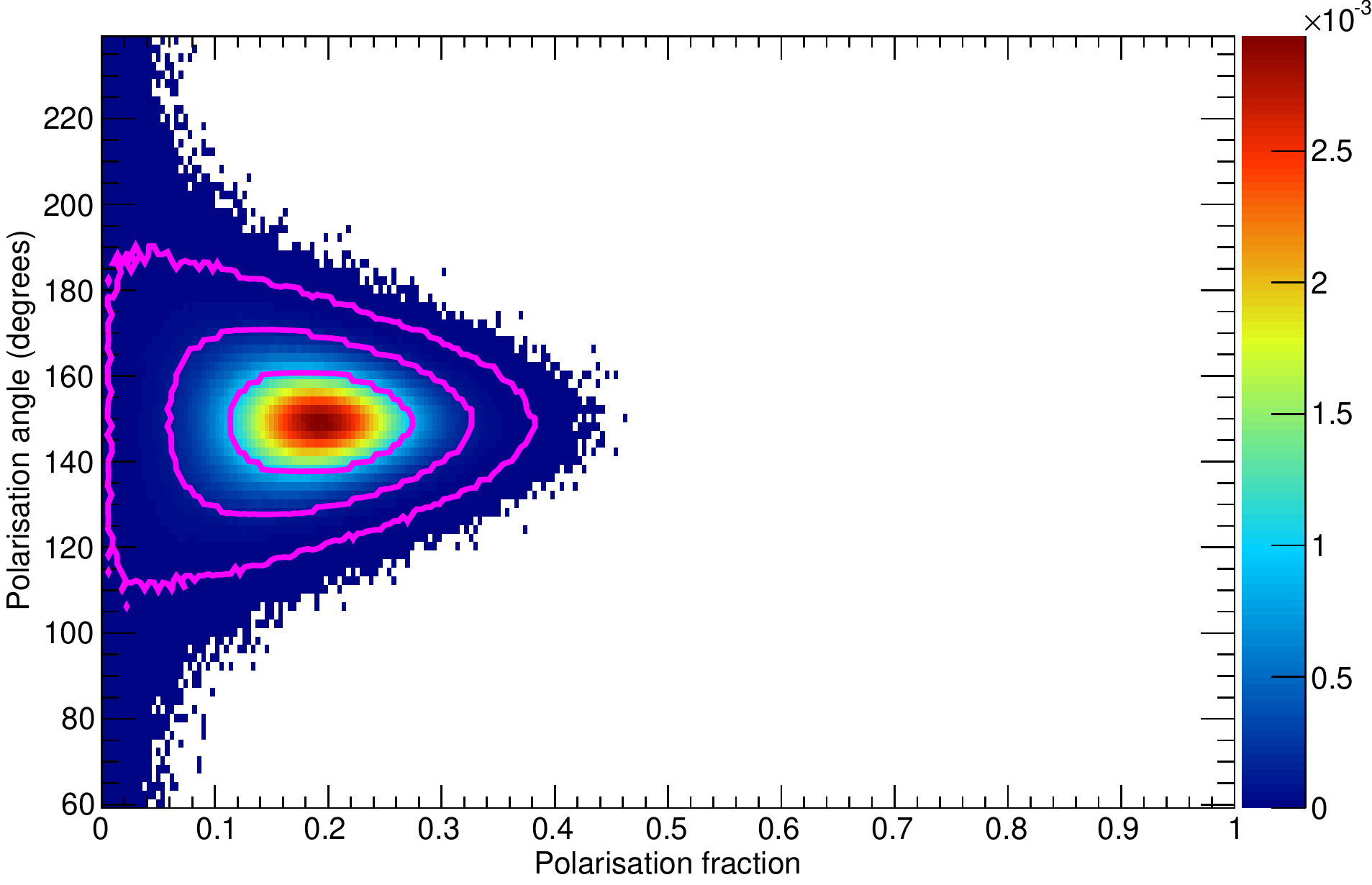}}
  [c]{\includegraphics[width=0.47\textwidth]{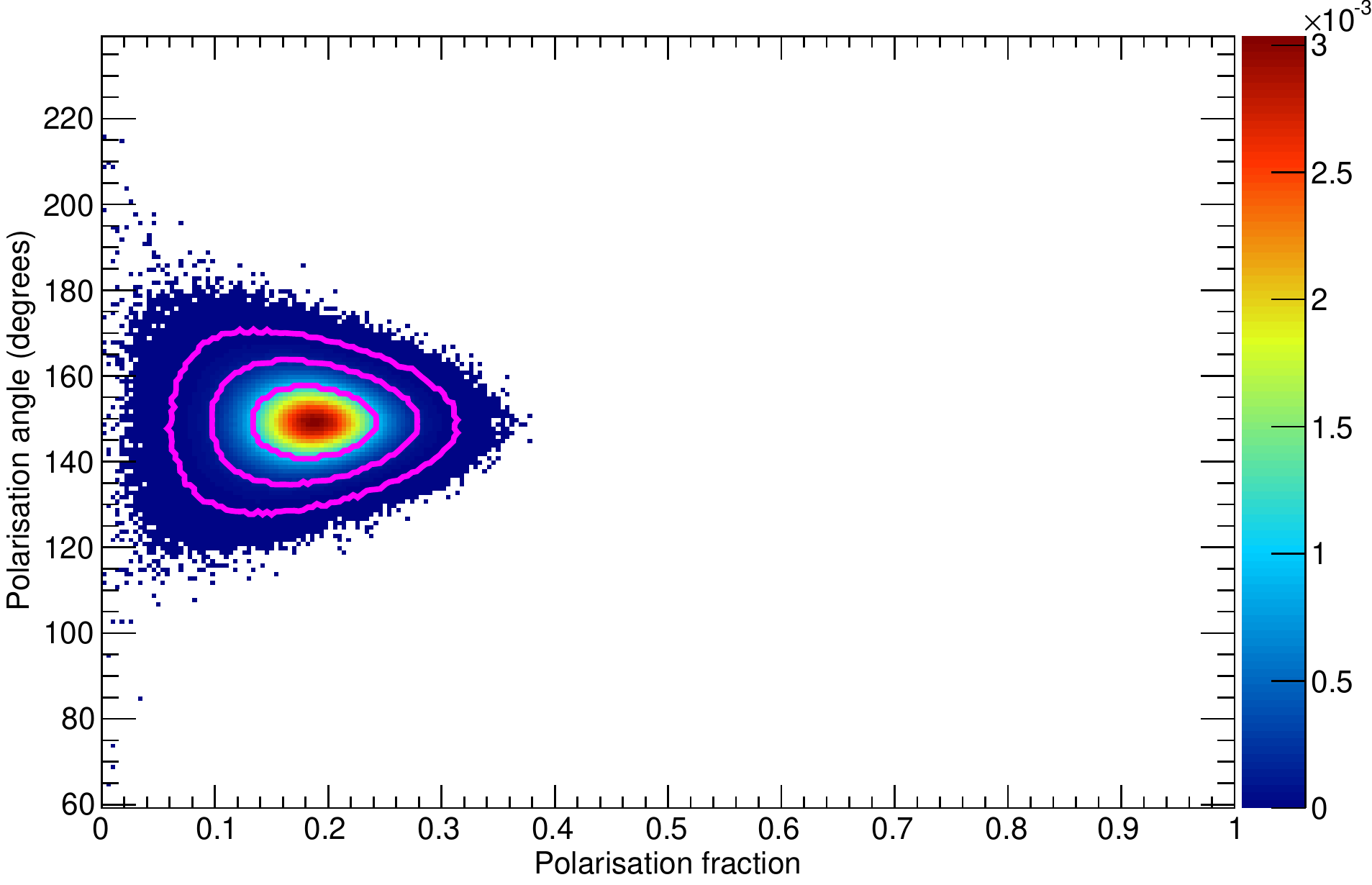}}\vspace{0.1cm}
  \caption{The posterior distribution of the polarisation fraction and polarisation angle of the Crab for: (a) the 2013 PoGOLite Pathfinder 
  observations~\citep{2015arXiv151102735C}, (b) PoGO+ same 7.3 hour observation time and (c) PoGO+ 15 hours observation time. The 
  two PoGO+ simulations assume the same central value of polarisation fraction and polarisation angle as (a). These results come from a 
  Bayesian analysis method developed by \cite{A&A_method}. The magenta lines are contours corresponding to 1, 2 and 3 standard 
  deviations Gaussian probability content. The colour bar is the probability per bin.}
  \label{Fig11}
\end{figure}
As an example of the quality of results that can be expected with such an increase in performance, a Crab observation with the 
reconstructed polarisation fraction and angle has been simulated using a Bayesian analysis method \citep{A&A_method}. This 
method uses the simulated $M_{100}$, Crab rate and background rate of PoGO+, as well as the expected polarisation fraction and 
polarisation angle as determined from the 2013 flight data, and produces a probability map of the reconstructed polarisation 
fraction and polarisation angle. 
Figure~\ref{Fig11} shows the results from the 2013 flight along with PoGO+ simulations for the same observation time as well as 
for a 5 day flight. 
While the PoGOLite Pathfinder measured a polarisation fraction of \mbox{(18.4$^{+9.8}_{-10.6}$)\%} \citep{2015arXiv151102735C}, 
PoGO+ would measure a polarisation fraction 3.4 standard deviations from zero at \mbox{(18.4$^{+5.32}_{-5.48}$)\%} for the 
same observation time of 7.3 hours and 5.0 standard deviations from zero at \mbox{(18.4$^{+3.72}_{-3.68}$)\%} for a 5 day 
flight. In addition, PoGO+ will be sensitive enough to separate the contribution from the nebula through pulse phase 
selections. Such selections would reduce the source rate by 65.5\% and the background rate by 60.0\% but assuming a relatively 
high degree of polarisation from the nebula\footnote{The polarisation fraction of the nebula is expected to be higher than the 
overall Crab emission 
\citep{2004ApJ...606.1125D}.} of 30\%, PoGO+ would make a measurement of 4.5 sigma significance with a 5 day observation 
(prediction made with the same method as Figure~\ref{Fig11}). Phase-resolved polarimetry of the Crab has never been achieved in 
the energy range of PoGO+ and will provide new information on the emission geometry and mechanisms at work.

The studies presented here conservatively assume the same waveform discrimination efficiency as in the 2013 flight. It should 
be noted that this efficiency is expected to increase in the final design due to the removal of the active collimators, which 
simplifies the waveform discrimination process significantly. The MDP of PoGO+ for the Crab after a 5 day flight with a 100\% 
waveform discrimination efficiency reaches $8.2\%$. The simulations also consider an average column density for the Crab and 
for \mbox{Cygnus X-1} observations, which is also conservative as it results in higher than average absorption by the atmosphere 
leading to an overestimate of the MDP by $\sim$10\% (relative).

It is also foreseen to apply software improvements for the next flight, for example by only storing two-hit events, the 
live-time increases by \mbox{$\sim$17\%} directly affecting the polarimeter performance. Increasing the live-time will allow 
in-flight trigger thresholds to be relaxed, increasing the low-energy sensitivity of PoGO+, which is especially important due 
to the power-law behaviour of the source spectra.

\section{Summary and Conclusions}
The PoGOLite Pathfinder has made a successful flight in 2013, completing an almost full circumnavigation of the North Pole. As 
a proof-of-concept mission originally intended to validate the polarimeter design and pointing system, 
the flight also resulted in a polarisation measurement from the Crab with a polarisation fraction of 
\mbox{(18.4$^{+9.8}_{-10.6}$)\%}. 
A thorough analysis of the flight data, the development of a 
detailed simulation, as well as a detailed and flight-validated background model have allowed the polarimeter design to be 
significantly improved in preparation for a re-flight. This upgraded design will improve the MDP by 44\% (relative), reaching 
a value of 11.1\% for the Crab and 10.5\% for \mbox{Cygnus X-1} (hard state) after a 5 day flight. Since these targets are 
visible at different times of the day, both can be studied in a single flight. Such performance will allow accurate 
polarisation measurement (5 standard deviation significance) of both sources in a previously unexplored energy range, 
20 - 240 keV. PoGO+ will also allow the measurement of the Crab nebula polarisation fraction through pulse phase selections, giving 
unique insight into the location and mechanisms of emission. 
The new design is currently being implemented at KTH Royal Institute of Technology in Stockholm, Sweden. The reassembly of the 
polarimeter is expected to be finished during the second half of 2015, allowing calibration tests to begin before the end of 
the year, in preparation for a re-flight from Esrange, Sweden, in the Summer of 2016.

\section*{Acknowledgments}
Funding received from The Swedish National Space Board, The Knut and Alice Wallenberg Foundation and The Swedish Research Council is gratefully acknowledged. 
KTH MSc students H\aa kan Wennl\"of and Philip Ekfeldt are thanked for their contributions to laboratory studies. All past members of the PoGOLite 
Collaboration not listed as authors on this paper are thanked for their important 
contributions to the development of the project.

\bibliographystyle{elsarticle-num}
\bibliography{biblist}

\end{document}